\documentclass[aps,pre,twocolumn]{revtex4}

\usepackage{graphicx}
\usepackage{subfigure}
	
\newcommand{\be}{\begin{equation}} 
\newcommand{\ee}{\end{equation}} 
\newcommand{\bea}{\begin{eqnarray}} 
\newcommand{\eea}{\end{eqnarray}} 
\newcommand{\bc}{\begin{center}} 
\newcommand{\ec}{\end{center}}

\begin{document}
\title{Simulation of Interdiffusion in Between Compartments Having
Heterogenously Distributed Donors and Acceptors}

\author{Erkan T\"uzel\footnote[2]{Present address: University of Minnesota, School of Physics and Astronomy,
116 Church St. SE, Minneapolis, MN, 55455, USA}}
\author{ K. Batuhan K{{\i}}sac{{\i}}ko{\u g}lu}
\author{{\"O}nder Pekcan}

\affiliation{Department of Physics, Faculty of  Sciences and Letters\\ 
Istanbul Technical University, Maslak 80626, Istanbul, Turkey}

\bibliographystyle{plain}
\pagenumbering{arabic} 

\begin{abstract}

The final stage of latex film formation was simulated by introducing donors and 
acceptors into the adjacent compartments of a cube. Homogenous and/or heterogeneous 
donor-acceptor distributions were chosen for different types of simulations. The 
interdiffusion of the donors and the acceptors within these cubes was generated 
using the Monte-Carlo technique. The decay of the donor intensity $I(t)$ by direct 
energy transfer (DET) was simulated for several interdiffusion steps. Gaussian noise 
was added to the $I(t)$ curves to obtain more realistic decay profiles. $I(t)$ decay 
curves were fitted to the phenomenological equation to calculate the fractional 
mixing at each interdiffusion step. The reliability of the Fickian diffusion model 
in the case of heterogenous and homogeneous donor-acceptor distributions are discussed 
for latex film formation. 

\end{abstract}
\maketitle

\section{Introduction} 

Polymer latex particles have been utilized in a wide variety of applications
in the coating and adhesive technologies, biomedical field, information
industry and microelectronics. In many of these applications, e.g., coatings
and adhesives, latexes form thin polymer films on a substrate surface.
Properties (mechanical, optical, transport, etc.) of the final film should
be tailor-made according to the application.

Film formation from latex particles is a complicated, multistage phenomenon
and depends strongly on the characteristics of colloidal particles. In
general, aqueous or non-aqueous dispersions of colloidal particles, with
glass transition temperature $(T_g)$ above the drying temperature, are named 
hard latex dispersion, however aqueous dispersion of colloidal particles
with $T_g$ below the drying temperature is called soft latex dispersion. 
The term ''latex film'' normally refers to a film formed from soft particles
where the forces accompanying the evaporation of water are sufficient to
compress and deform the particles into a transparent, void-free
film$^{1,2}$. However, hard latex particles remain essentially discrete and
undeformed during drying process. Film formation from these dispersion can
occur in several stages. In both cases, the first stage corresponds to the
wet initial state. Evaporation of solvent leads to second stage in which the
particles form a closed pack array, here if the particles are soft they are
deformed to polyhedrans (see Figure 1). Hard latex however stay undeformed at 
this stage. Annealing of soft particles cause diffusion across particle-particle
boundaries which leads the film to a homogeneous continuous material.
In the annealing of hard latex system, however deformation of particles first leads
to void closure$^{3,4}$ and then after the voids disappear, diffusion across
particle-particle boundaries starts, i.e. the mechanical properties of hard
latex films can be evolved by annealing; after all solvent has evaporated
and all voids have disappeared.

\begin{figure}
\bc
\leavevmode
\includegraphics[width=6cm,angle=0]{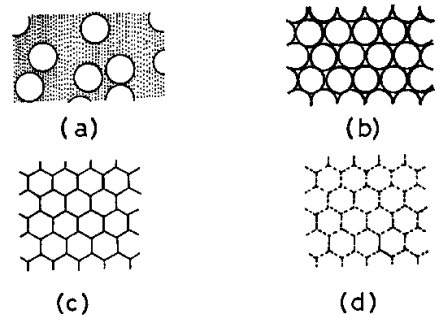}
\caption{\small{
A pictorial representation of the stages of latex film formation from soft polymer 
particles. a) The latex dispersion. b) The solvent evaporates leaving the particles in close contact.
c) Deformation and packing of the particles. d) Further coalescence produces a mechanically rigid film.
}}
\ec
\end{figure}

Transmission electron microscopy (TEM) has been used to examine the
morphology of dried latex films$^{5,6}$. These studies have shown that in
some instances the particle boundaries disappeared over time, but in other
cases the boundaries persisted for months. It was suggested that in the
former case particle boundaries were healed by polymer diffusion across the
junction. In the last few years, it has become possible to study latex film
formation at the molecular level. Small-angle neutron scattering (SANS) was
used to examine deuterated particles in a protonated matrix. It was observed
that the radius of the deuterated particle increased in time as the film was
annealed$^7$ and as the polymer molecules diffused out of the space to which
they were originally confined. The process of interparticle polymer
diffusion has been studied by the direct energy transfer (DET) method, using
transient fluorescence (TRF) measurements$^{8,9}$ in conjunction with latex
particles labelled with donor and acceptor chromophores. The steady state
fluorescence (SSF) method combined with DET was also used for studying film
formation from hard latex particles$^{10-13}$. An extensive review of the
subject is given in reference 14. In DET measurements distribution of donors
and acceptors are thought to be crucial i.e. it is believed that donors and
acceptors have to be distributed randomly in the latex particles for the
reliable TRF measurements, to determine the diffusion coefficients, D of
polymer chains. 

Recently we have performed various experiments with photon transmission method 
using an U.V. Visible (UVV) spectrophotometer to study latex film
formation from PMMA and PS latexes in where void closure and interdiffusion
processes at the junction surfaces are studied$^{15-18}$. All these studies 
indicate that annealing leads to polymer diffusion and mixing as the
particle junction heals during latex film formation. Recently, Monte Carlo 
simulation of interdiffusion and its monitoring by DET during latex film formation 
has also been studied in our laboratories$^{19,20}$. 

In this work, Monte Carlo method was used to simulate the final stage of film
formation by introducing donors and acceptors into the adjacent compartments
of a cube. Four different combinations of donor-acceptor distributions were
chosen for the different types of simulations. For example in the first case
distribution of donors and acceptors in their adjacent compartments are taken 
as homogenous and gaussian respectively. In the second case distributions
are switched from one compartment to the other. In the third case, both
distributions are taken as gaussian and in the final case, distribution of donors
and acceptors are both taken homogeneously to compare this case with the others. 

The interdiffusion of donors and acceptors between these adjacent
compartments was randomly generated by Monte Carlo method. The decay of the
donor intensity, $I(t)$ by DET was simulated for several interdiffusion
steps and a gaussian noise was added to generate the realistic time resolved
fluorescence data. I(t) decays were fitted to the phenomenological equation
to obtain the fractional mixing at each interdiffusion step. The
reliability of the Fickian model and the effect of heterogenous
donor-acceptor distributions are discussed at the last stage of latex film
formation process.  

\section{DET and Fluorescence Decay}

\begin{figure}[t]
\bc
\leavevmode
\includegraphics[width=5cm,angle=0]{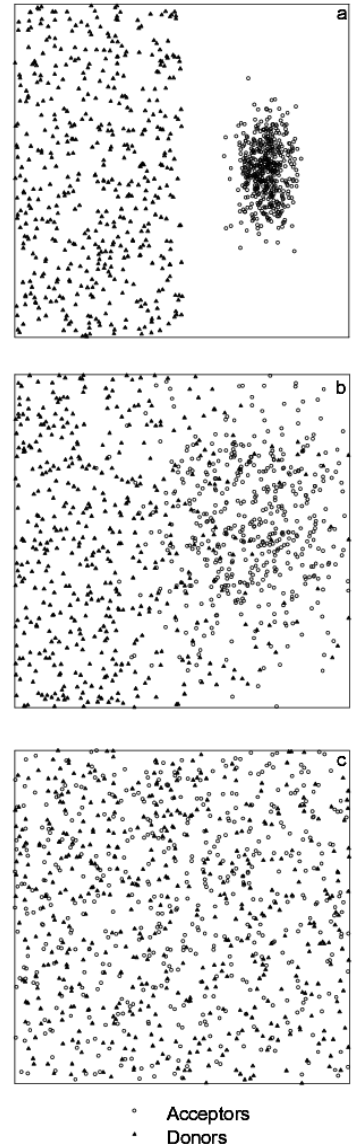}
\caption{\small{
Several snapshots of the interdiffusion process between adjacent compartments of a cube 
in which donors and acceptors are distributed homogenous and gaussian wise. 
a) K=0.0, b) K=0.3 and c) K=1.0 $\;$.
}}
\ec
\end{figure}

\begin{figure}[t]
\bc
\leavevmode
\includegraphics[width=5cm,angle=0]{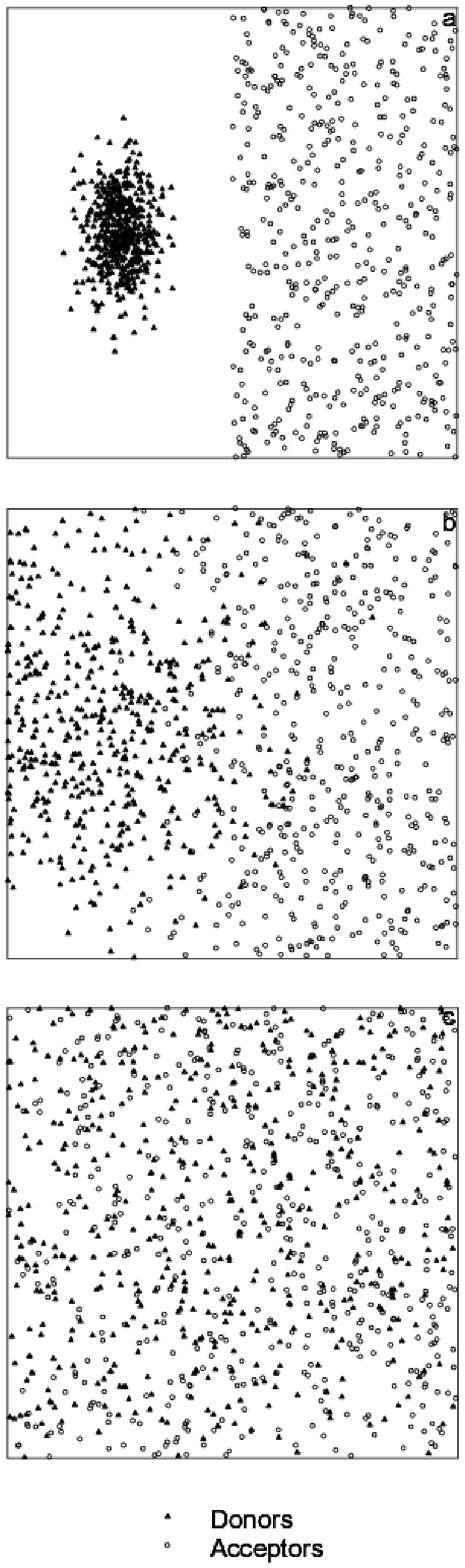}
\caption{\small{
Several snapshots of the interdiffusion process between adjacent compartments of a cube 
in which donors and acceptors are distributed gaussian and homogenous wise. 
a) K=0.0, b) K=0.3 and c) K=1.0 $\;$.
}}
\ec
\end{figure}

\begin{figure}
\bc
\leavevmode
\includegraphics[width=5cm,angle=0]{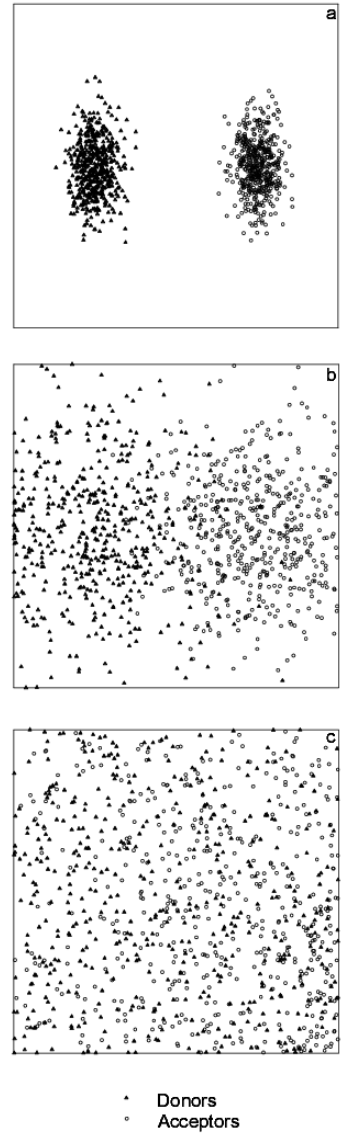}
\caption{\small{
Several snapshots of the interdiffusion process between adjacent compartments of a cube 
in which both donors and acceptors are distributed gaussian wise. 
a) K=0.0, b) K=0.3 and c) K=1.0 $\;$.
}}
\ec
\end{figure}

\begin{figure}
\bc
\leavevmode
\includegraphics[width=5cm,angle=0]{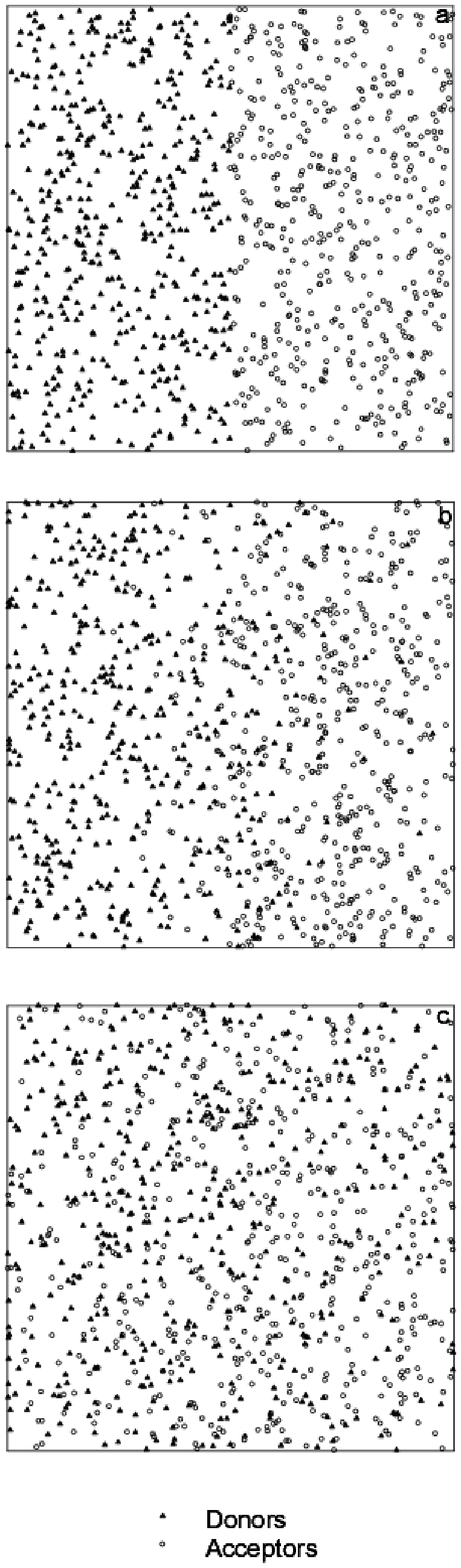}
\caption{\small{
Several snapshots of the interdiffusion process between adjacent compartments of a cube 
in which both donors and acceptors are distributed homogenously. 
a) K=0.1, b) K=0.4 and c) K=1.0 $\;$.
}}
\ec
\end{figure}

Polymer diffusion obeys de Gennes scaling laws for times short compared 
to the tube renewal time $t_{tr}$, but for long times it is like a random 
walk process (Fickian diffusion). In order to be able to determine whether 
the diffusion is Fickian, one must compare the experimental data with the 
results of simulations of DET with Fickian diffusion. 
 
TRF in conjunction with the DET method, monitors the extent of interdiffusion 
of donor (D) and the acceptor (A) labelled polymer molecules. The sample is 
made of a mixture of D and A labelled latex spheres. When this sample is 
annealed for a period of time and the donor fluorescence profiles are 
measured, each decay trace provides a snapshot of the extent of 
interdiffusion$^{9}$. A film sample after annealing was considered to be 
composed of three regions; unmixed D, unmixed A and the mixed D - A region. 
This model was first empirically introduced by the two component donor 
flourescence decay$^{21,22}$. 

When donor dyes are excited using a very narrow pulse of light, the excited 
donor returns to the ground state either by emitting a fluorescence photon or 
through the nonradiative mechanism. For a well behaved system, after exposing 
the donors with a short pulse of light, the fluorescence intensity decays 
exponentially with time. However, if acceptors are present in the vicinity of 
the excited donor, then there is a possibility of DET from the excited donor 
to the ground state acceptors. In the classical problem of DET, neglecting 
back transfer, the probability of the decay of the donor at $r_k$ due to the 
presence of an acceptor at $r_i$ is given by$^{23}$ 
 
\be 
P_k(t) = exp[-t/{\tau_0}-{w_{ik}}t] 
\ee 

\noindent 
where $w_{ik}$ is the rate of energy transfer given by F{\"o}rster as 
 
\be 
w_{ik} = {\frac {3}{2}}{{\kappa}^2}{\frac{1}{{\tau}_0}}{({\frac{R_0}{r_{ik}}})}^6 
\ee 
 
\noindent
Here $R_0$ represents the critical F{\"o}rster distance and $\kappa$ is a 
dimensionless parameter related to the geometry of interacting dipole. 
If the system contains $N_D$ donors and $N_A$ acceptors, then the donor 
fluorescence intensity decay can be derived from the equation (2) and given 
by$^{16}$ 
 
\bea
\frac{I(t)}{I(0)}&=&\exp(-t/{\tau_0})\frac{1}{N_D}\int{{n_D} 
(r_k)dr_k} \nonumber \\
&&\times \prod_{i=1}^{N_A} {\frac{1}{N_A}}\int{{n_A}{(r_i)}}dr_i\exp{(-w_{ik}t)} 
\eea 

\noindent
Here $n_D$ and $n_A$ represent the distribution functions of donors and 
acceptors. In the thermodynamic limit equation (3) becomes$^{16}$ 
 
\bea
\frac{I(t)}{I(0)}&=&\exp(-t/{\tau_0})\frac{1}{N_D}\int{{n_D}(r_k)dr_k} \nonumber \\
&&\times \exp({-\int{n_A(r_i)dr_i(1-\exp{(-{w_{ik}}t))}}}) 
\eea 

\noindent
This equation can be used to generate donor decay profiles by Monte-Carlo 
techniques. It is shown that the equation (4) reduces to a more simple
form which can be compared to the experimental data$^{3}$. Their argument is 
summarized below for clarity. Changing to the coordinate $r_{ik}=r_i-r_k$ 
leads to, 
 
\bea
\frac{I(t)}{I(0)}&=&\exp(-t/{\tau_0})\frac{1}{N_D}\int{{n_D}(r_k)dr_k} \nonumber \\
&\times& \prod_{i=1}^{N_A} \int_{r_K}^{R_g-r_K}{{n_A}{({r_{ik}}+r_k)}}dr_{ik}
\exp{(-w_{ik}t)} 
\eea	

\noindent 
where $R_g$ is an arbitrary upper limit. Placing a particular donor at the 
origin and assuming that the mixed and unmixed regions are created during 
interdiffusion of D and A, the equation (5) becomes 
 
\bea 
\frac{I(t)}{I(0)}&=&B_1\exp(-t/{\tau_0})\prod_{i=1}^{N_A}\frac{1}{N_A}
\int_{0}^{R_g}{{n_A}(r_{ik})dr_{ik}} \nonumber \\
&&\times \exp{({-w_{ik}}t)}+B_2 \exp(-t/{\tau_0}) 
\eea

\noindent 
where 
 
\be 
B_{1,2}=\frac{1}{N_D}\int_{1,2}{n_D(r_k)dr(k)} 
\ee 
 
\noindent 
represent the fraction of donors in mixed and unmixed regions, respectively. The 
integral in equation (6) produces a F{\"o}rster type of function$^{24,25}$ 
 
\be
\prod_{i=1}^{N_A} \frac{1}{N_A} \int_0^{R_g} n_A(r_{ik})dr_{ik}exp(-w_{ik}t)=
exp(-C({\frac{t}{\tau_0}})^{1/2})
\ee
 
\noindent 
where C is proportional to acceptor concentration. Eventually, one gets the 
following formula for the fluorescence intensity.

\be 
\frac{I(t)}{I(0)}=B_1\exp(-t/{\tau_0}-C(\frac{t}{\tau_0})^{1/2})+
B_2\exp(-t/{\tau_0}) 
\ee 

\noindent 
Here it is useful to define the mixing ratio K representing the order of mixing 
during interdiffusion of the donors and the acceptors as 
 
\be 
K=\frac{B_1}{B_1+B_2}
\ee 
 
\section{Simulation of Interdiffusion} 
 
The interdiffusion of donors and acceptors between two adjacent compartments 
corresponds to the last stage of latex film formation process. Here the geometry 
is simplified using cubes instead of the polyhedrons, and donors and acceptors are 
randomly distributed in seperate adjacent compartments in a cube. Figure 2a, 3a, 
4a and 5a present the four types of combinations of donor-acceptor distributions. 
In Figure 2a donors and acceptors are distributed in the adjacent compartments in 
homogeneous and gaussian wise distributions, respectively. When these distributions 
are completely inversed, the situation is presented in Figure 3a. Figures 4a and 5a 
present acceptors and donors both distributed in separate compartments in gaussian 
and in homogeneous wise distributions, respectively. 

\begin{figure}
\bc
\leavevmode
\includegraphics[width=7cm,angle=0]{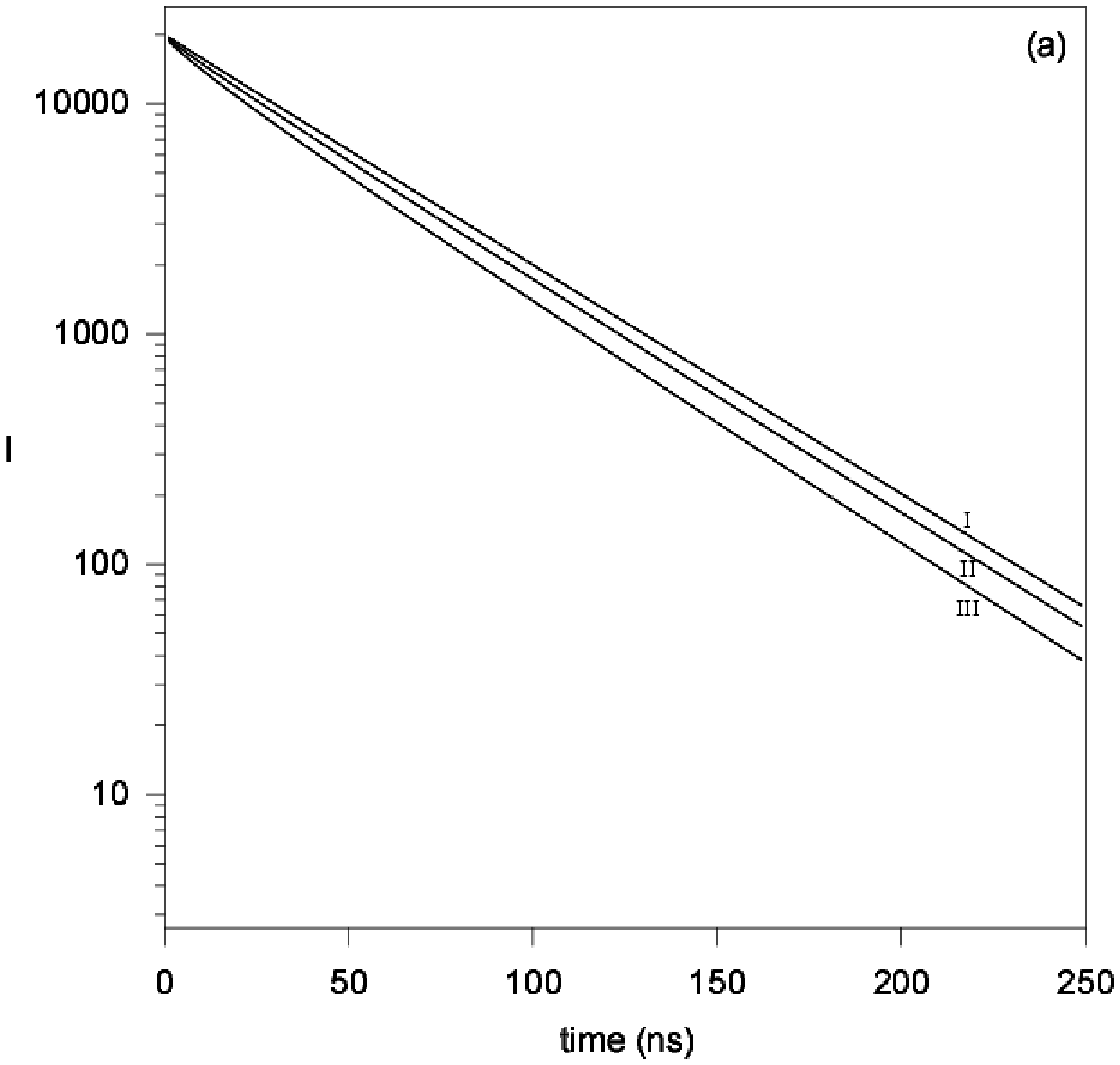}
\bigskip
\includegraphics[width=7cm,angle=0]{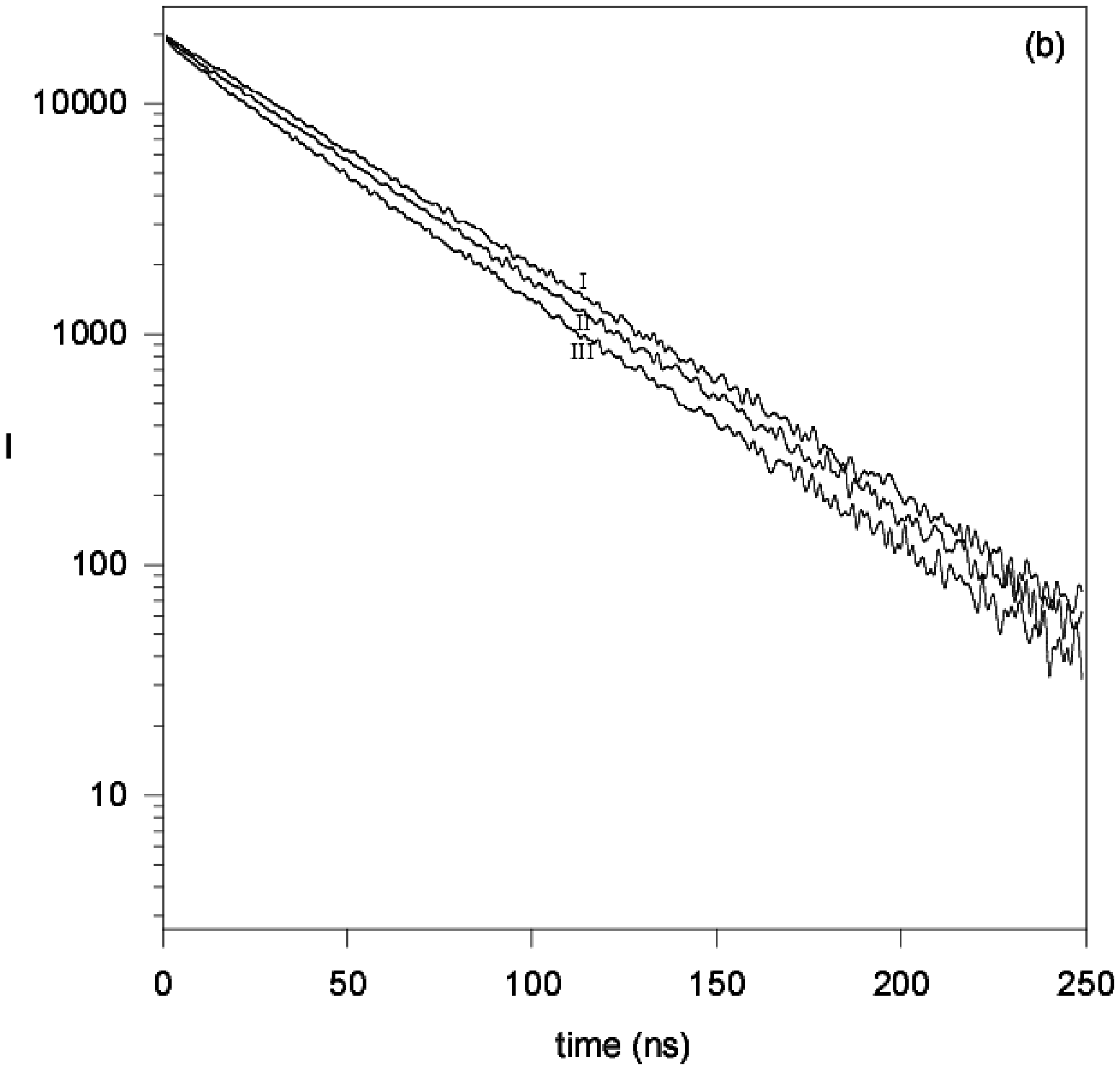}
\caption{\small{
a) Decay curves at the several interdiffusion steps for both donors and acceptors are homogenous 
distributed in adjacent compartments. 
I) K=0.1, II) K=0.5, III) K=1.0
b) Noisy decay curves for the above picture. 
}}
\ec
\end{figure}

\begin{figure}
\bc
\leavevmode
\includegraphics[width=7cm,angle=0]{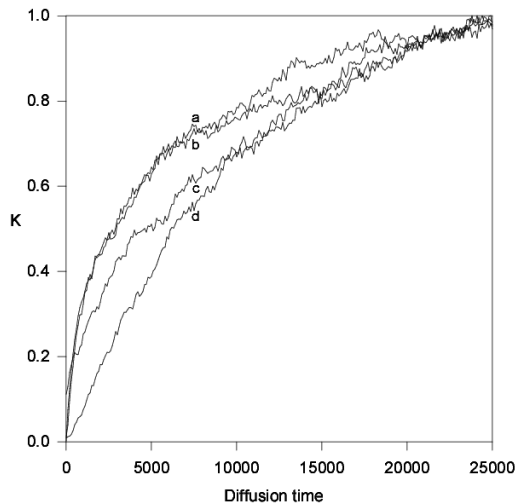}
\caption{\small{
Comparison of the plots of the mixing ratio K versus diffusion time for different initial 
distributions as a) Donors and acceptors are distributed homogenous and gaussian wise, 
b) Donors and acceptors are distributed gaussian and homogenous wise, 
c) Both donors and acceptors are distributed gaussian wise, d) Both donors and acceptors are distributed homogenously.
}}
\ec
\end{figure}

\begin{figure*}[!]
\bc

\subfigure[]{\includegraphics[width=7cm,angle=0]{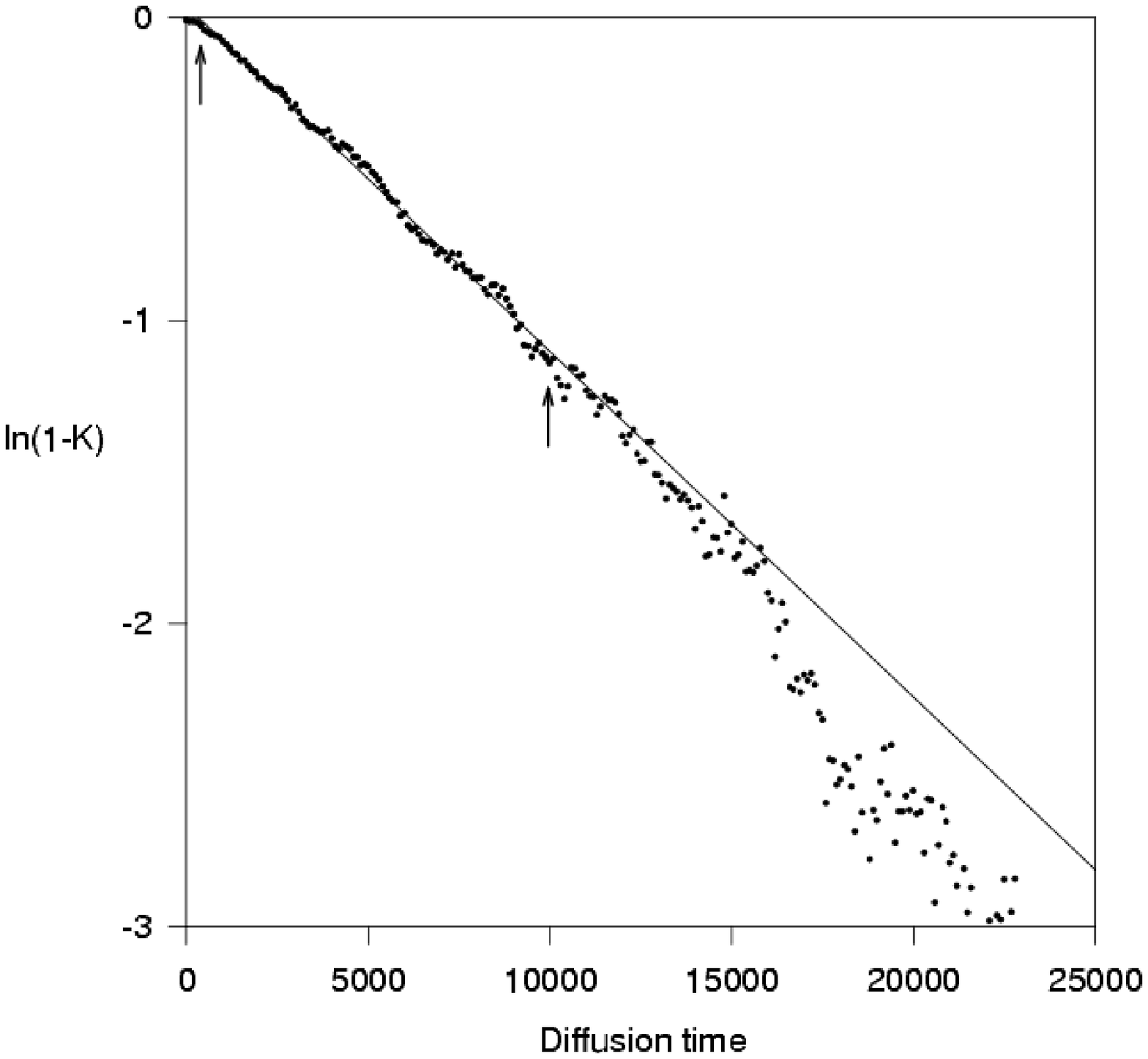}}\hspace{.25in}
\subfigure[]{\includegraphics[width=7cm,angle=0]{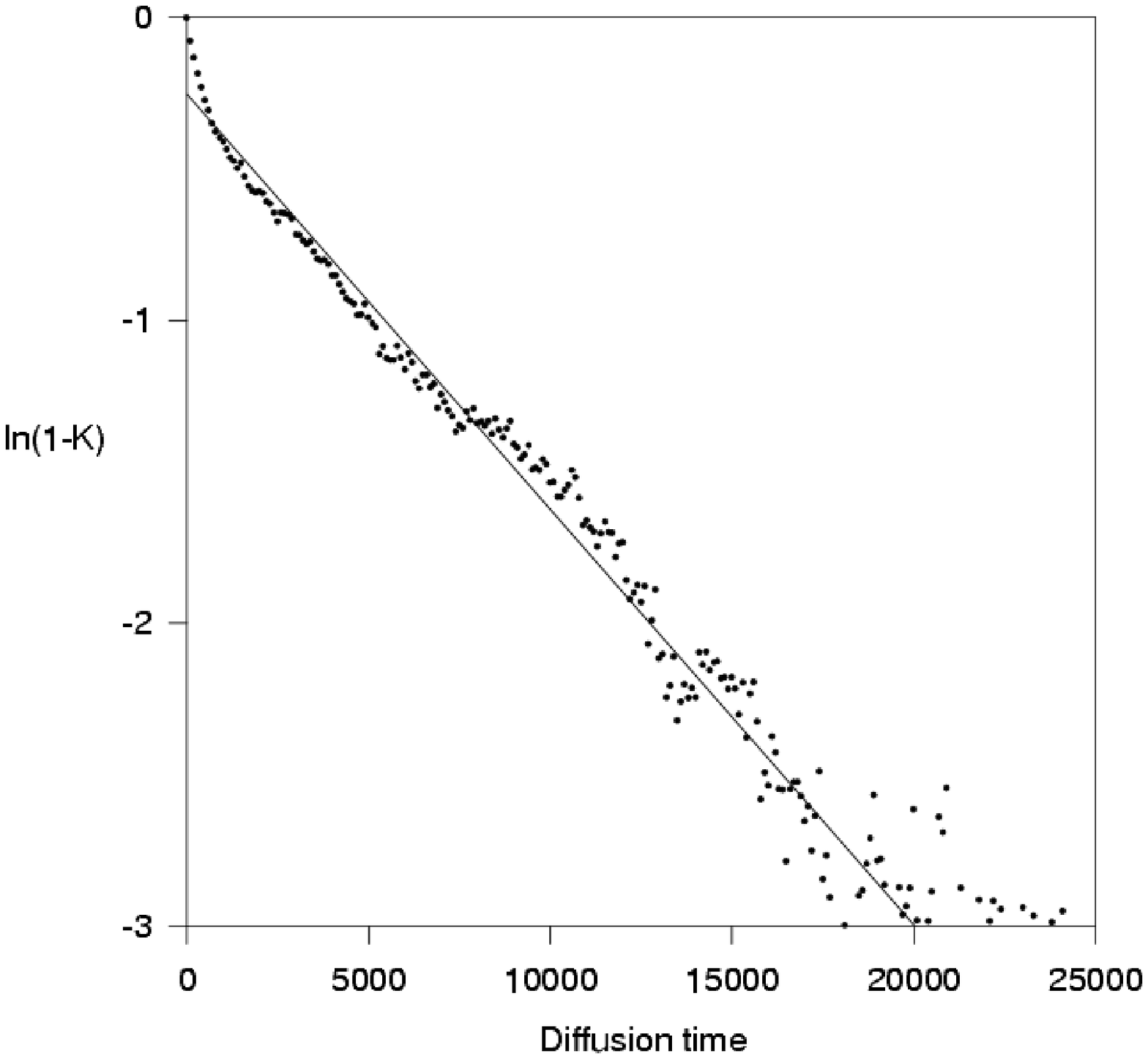}}
\subfigure[]{\includegraphics[width=7cm,angle=0]{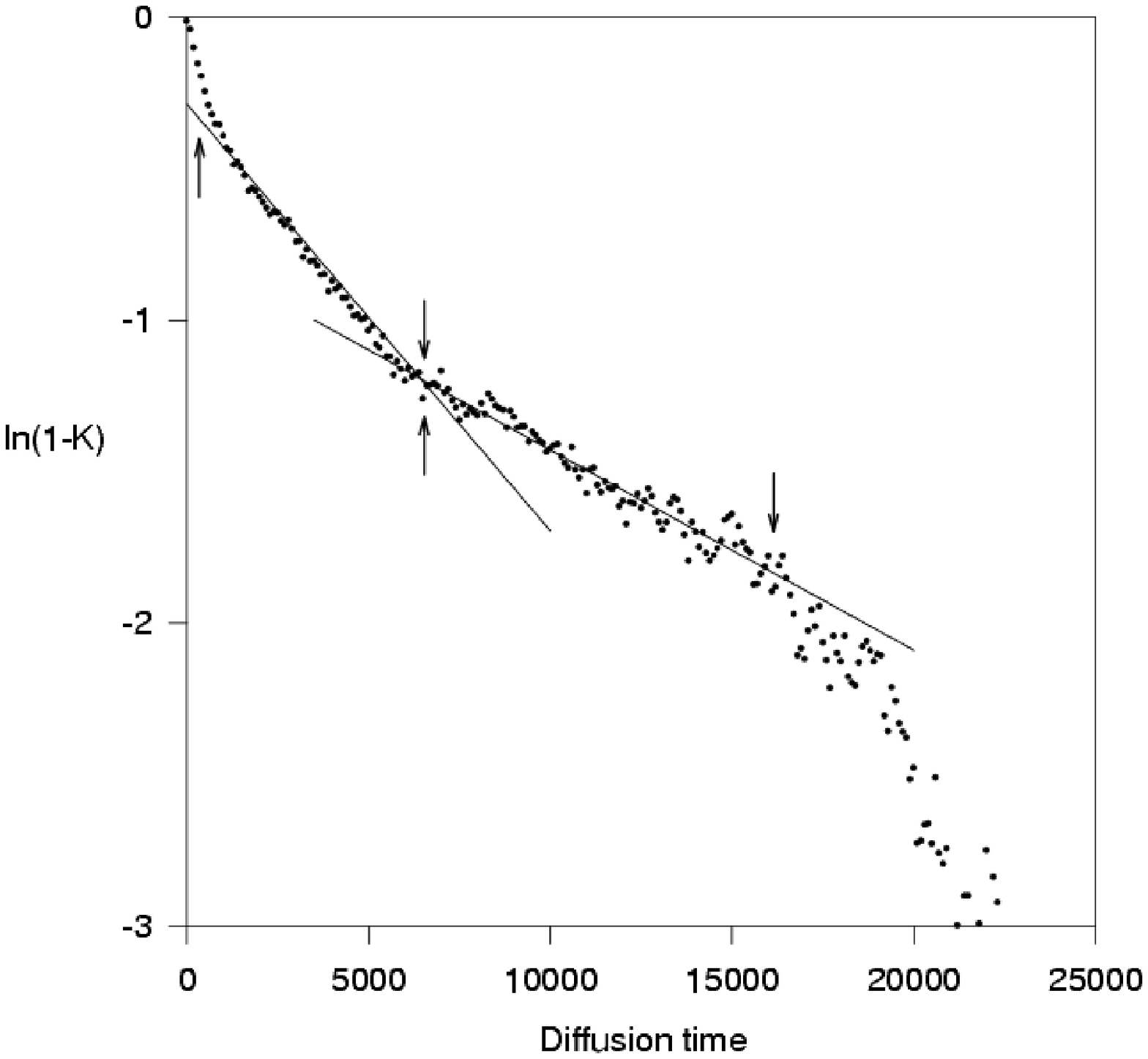}}\hspace{.25in}
\subfigure[]{\includegraphics[width=7cm,angle=0]{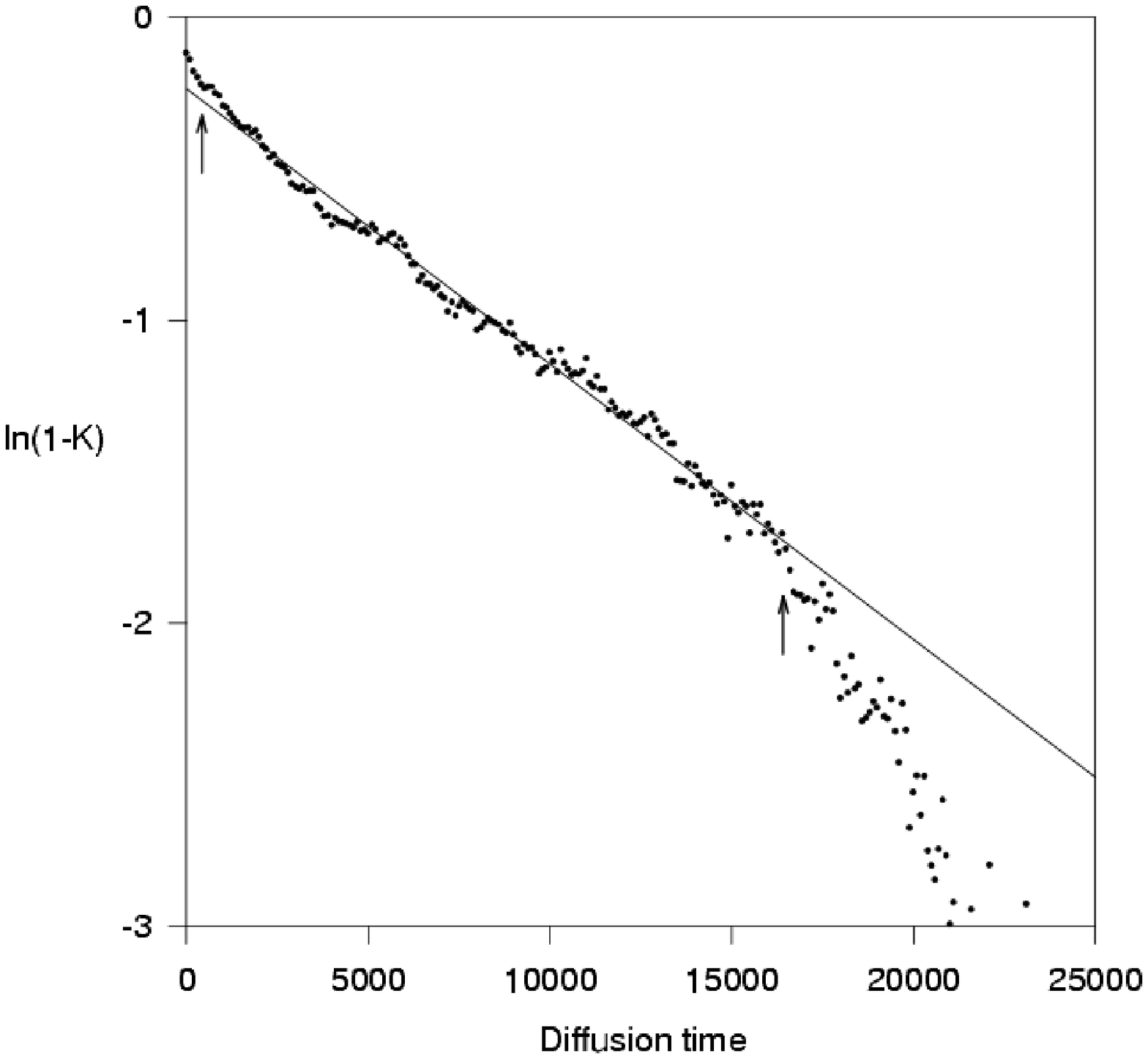}}

\caption{\small{The plots of $\ln (1-K)$ versus diffusion time obtained for different combinations of distributions
given in Figure 7. The solid lines present the fit of the data to equation (19). The slope of the 
solid lines produced diffusion constants which are listed in Table I. The regions used for the linear fits are shown within arrows in Figures 8a, 8c and 8d. In Figure 8b all the data points are
used in the fit. }}
\ec
\end{figure*}

Figures 2b, 3b, 4b and 5b present the picture after the Brownian motion of donors 
and acceptors generated for several interdiffusion steps for each combination of 
donor-acceptor pairs which are given in Figures 2a, 3a, 4a and 5a, respectively. 
In each diffusion step, all the donors and acceptors move within a range 
of 0 to 1 $A^o$ in any direction, but are reflected from the sides of the cube. After 
each diffusion step, the diffusion time increments one unit. $ 25 \times 10^3 $ diffusion 
steps were used for all sample simulations. The decay of donor intensity by DET is 
simulated for the configurations at the end of each 100 step of diffusion, therefore 
the diffusion process can be monitored quite clearly and accurately. Moreover, the
average is taken over 10 different runs for each initial distribution. Figures 2c, 3c,  
4c and 5c present the final picture of the interdiffusion between two adjacent compartments 
in a cube.  

The donor decay profiles were generated using equation (4). The side of the cube, L, is 
taken as 500 $A^o$ and the F{\"o}rster distance as 26 $A^o$. The number of donors, $N_D$, 
and acceptors $N_A$, are both chosen as 500. The $w_{ik}$  values for each donor-acceptor 
pair are obtained from equation (2). The parameter $\kappa^2$ is chosen as 0.476, a value 
appropriate for immobile dyes$^{20}$, and the donor lifetime $\tau_d$ is taken as 44ns. 
Equation (4) is then used to simulate the donor intensity I(t) decay profiles. 
$I(0)=2 \times 10^4$ is chosen and the decay profiles are obtained for a 250ns interval, 
divided into 250 channels of 1ns each. Decay profiles at the several interdiffusion steps 
for both donors and acceptors are homogenously distributed in adjacent compartments are 
presented in Figure 6a.  

Here, one may also take into account the effect of the lamp profile when calculating 
the decay profiles$^{19,20}$. To do so the decay profiles generated by the Monte Carlo 
simulation should be convolved with an experimental lamp profile, then the experimentally 
measured $\phi(t)$ is obtained by convolution of $I(t)$ with the instrument response 
function $L(t)$, as 

\be 
\phi(t)=\int_{0}^{t} L(t)I(t-s)ds 
\ee

\noindent
In this work, since we are interested in the effect of donor-acceptor 
distributions on the interdiffusion, instead of using experimental decay 
profiles we used generated decay profiles. This assumption is valid if one 
uses a delta, $\delta$ function light source (e.g. a very fast laser) as the 
lamp profile. In this case no convolution is needed and equation (11) produces
$I(t)$. However, to obtain more realistic decay profiles, gaussian noise can be
added to the original decay profiles using Box, Muller and Marsaglia$^{24}$ 
algorithm. In this algorithm, at first two gaussian numbers ( $ U_1 $ , $ U_2 $ ) 
between $0$ and $1$ are created. Then $ V_1 $ and $ V_2 $ are calculated as 
shown below 
 
\bea 
V_1 = 2U_1 - 1 \\ 
V_2 = 2U_2 - 1 
\eea 
 
\noindent 
Both $ V_1 $ and $ V_2 $ are distributed randomly in the range $[-1,1]$. S is 
calculated from these two numbers. 
 
\be 
S = {{V_1}^2} + {{V_2}^2} 
\ee 
 
\noindent 
If $S < 1$, operation is unsuccessful and new $ U_1 $ and $ U_2 $ numbers 
are created. If $S > 1$, $ X_1 $ and $ X_2 $ are calculated as shown below. 
 
\bea 
M & = & q (\frac{-2\ln S}{S})^{\frac {1}{2}} \\ 
X_1 & = & (V_1 M)+p \\ 
X_2 & = & (V_2 M)+p 
\eea 
 
\noindent 
$ X_1 $ and $ X_2 $ are mutually independent. They are gaussian numbers with an average 
p and standard deviation q. The noisy decay profiles for the homogenously distributed 
donors and acceptors at several interdiffusion steps are shown in Figure 6b. 

In order to calculate the mixing ratio, K defined in equation (10) one should fit the 
generated decay profiles to equation (9). The decay profiles were fitted to equation (9) 
using Levenberg-Marquart$^{25}$ algorithm. During fits the parameters C and $\tau_0$ are 
kept constant (C=1) and only the parameters $B_1$ and $B_2$ are varied. More than $5000$
decay profiles are fitted and the goodness of fitting is varied around $\chi^2 < 1.5$. The 
produced $B_1$ and $B_2$ values are used to obtain K values at each interdiffusion step. 
Figure 7 compares the plots of K versus diffusion time for the interdiffusions presented 
in Figures 2, 3, 4 and 5. Each curve in Figure 7 is obtained from the average of a set of 
10 runs. 
 
To test whether the simulated interdiffusion is Fickian or not, the planar sheet model is  
chosen$^{26}$. In this model the fraction of the diffusing substance that has diffused out 
of the planar sheet at time t is given by 
 
\be 
K_s = \frac {8}{\pi} \sum_{n=0} \frac {1}{{(2n+1)}^2} \exp{(-\frac {D{(2n+1)}^2 \pi^2 t}{a^2})} 
\ee 
 
\noindent 
where D is the diffusion constant and a is the maximum distance over which 
diffusion can occur. Since $\lim_{k\rightarrow \infty} K_s = 1$, eq.(18) can be 
written for $n=0$ in the form 
 
\be 
\ln (1-K_s) = -\frac {D\pi^2 t}{a^2} 
\ee 
 
\noindent 
$\ln (1-K) $ values are plotted versus diffusion time in Figure 8 and were fitted to 
equation (19). The fits obtained for all of the four combinations of distributions are 
shown in Figures 8a, 8b, 8c and 8d. The solid lines in the plots represent the 
fitting curve and the dots represent the digitized data. The diffusion constants and the 
correlation coefficients showing the goodness of fits are presented in Table I. 

\begin{table}
\caption{ $D\pi^2/a^2$ values are produced by fitting the data in Figure 7 to the equation (19). 
The fits are presented in Figure 8 for the various combinations of distributions. 
$R^2$ is the correlation coefficient for the fits. }

\begin{ruledtabular} 
\begin{tabular}{cccc}
{\bf Donor} & {\bf Acceptor} & ${\bf D\pi^2/a^2 (\times 10^{-4})}$ & ${\bf R^2}$           \\ \hline
Homogenous  & Gaussian       & 1.14 $\pm$ 0.01                             & 0.995               \\
Gaussian    & Homogenous     & 1.37 $\pm$ 0.03                             & 0.925               \\
Gaussian    & Gaussian       & 1.41 $\pm$ 0.03                             & 0.973   \\
	    &		     & 0.66 $\pm$ 0.02                             & 0.958   \\
Homogenous  & Homogenous     & 0.91 $\pm$ 0.01                             & 0.991               
\end{tabular}
\end{ruledtabular}
\end{table}

\section{Conclusions} 

Fits in Figure 8 and the values in Table I strongly suggest that people who work in 
TRF area have to be very careful to synthesize their latex particles which are labelled
with the fluorescence dyes. In this work, it is observed that when the dye distribution 
is not homogenous different results in interdiffusion processes can be produced even the 
latex particles are in equal size. All data in Figure 8 present that interdiffusion 
saturates at the long time region. At the short time region the initial donor-acceptor 
distribution is quite critical and effects the interdiffusion (mixing ratio, K). When 
donors are distributed in gaussian wise, delay for the onset of interdiffusion is observed
at early time region which is obvious, since it takes some time for the donors to reach
the acceptors to perform DET. In this case if the acceptors are distributed homogenously, 
interdiffusion occurs with a single diffusion constant D, however if the acceptors are
distributed gaussian wise, two different interdiffusion regimes can be observed at 
the intermediate time region. In other words, after a certain delay at early times, donors 
and acceptors meet quite fast to perform DET and then interdiffusion slows down and 
finally mixing saturates at longer times. 

When the donors are distributed homogenously the delay at the short time region is quite
small, especially if the acceptors are distributed gaussian wise, no delay is observed. 
In this case when the acceptors are distributed either gaussian or homogenous wise, single 
interdiffusion regime is observed at intermediate time region where in both cases 
interdiffusion rate is similar and much smaller than when the donors are distributed 
gaussian wise (see Table I). 

In conclusion, if one assumes that the ideal distribution for donors and acceptors in latex
particles are both homogenous, then one has to expect that experimental results for K should
obey the picture in Figure 8d, even though the picture in Figure 8a looks much better i.e. 
interdiffusion starts with no delay and produces single interdiffusion constant. 
 
\section*{Acknowledgements} 
We would like to thank Professor A. T. Giz for his critical comments and discussions.

\end{document}